\documentclass[11pt]{article}
\usepackage{graphicx} 

\usepackage{amssymb}
\usepackage[margin=1.165in]{geometry}
\usepackage{amsfonts}
\usepackage{ragged2e}
\usepackage{multirow}
\usepackage{amsmath}
\usepackage{physics}
\usepackage{enumitem}
\usepackage{textcomp}
\usepackage{setspace}
\usepackage{tabularx}
\usepackage{datetime2}
\usepackage{authblk}
\usepackage[dvipsnames]{xcolor}
\usepackage{bm}
\usepackage{orcidlink}
\usepackage{soul}
\usepackage{tcolorbox}

\usepackage[
backend=biber,
style=numeric-comp,
sorting=none
]{biblatex}
\usepackage{chngcntr}
\usepackage{tikz}
\usepackage{comment}

\usepackage{hyperref}
\hypersetup
{
colorlinks=true,
linkcolor=blue,
urlcolor={winered},
filecolor={winered},
citecolor={winered},
allcolors={blue}
}

\definecolor{plotgreen}{RGB}{20, 164, 28}
\definecolor{plotred}{RGB}{195, 7, 24}

\definecolor{desred}{RGB}{231,101,109}
\definecolor{desblue}{RGB}{105,105,253}
\definecolor{desgreen}{RGB}{120, 213, 124}
\definecolor{despurp}{RGB}{163, 73, 164}

\newcommand{\tikzcircle}[2][red,fill=red]{\tikz[baseline=-0.5ex]\draw[#1,radius=#2] (0,0) circle ;}%

\title{\bf{Effective Metric Description \\ \ of \\  2+1 Dimensional Quantum Black Holes}}

\date{}
\author[1]{Stefan Hohenegger\footnote{{\Large\orcidlink{0000-0001-6564-0795}}\hspace{0.1cm}\raisebox{0.9ex}{\href{mailto:s.hohenegger@ipnl.in2p3.fr}{s.hohenegger@ipnl.in2p3.fr}}}}
\author[2]{Mikolaj Myszkowski\footnote{{\Large\orcidlink{0000-0002-5207-4509}}\hspace{0.1cm}\raisebox{0.9ex}{\href{mailto:mikolaj@qtc.sdu.dk}{mikolaj@qtc.sdu.dk}}}}
\author[2]{Mattia Damia Paciarini\footnote{{\Large\orcidlink{0009-0000-1044-341X}}\hspace{0.1cm}\raisebox{0.9ex}{\href{mailto:damiapaciarinim@qtc.sdu.dk}{damiapaciarinim@qtc.sdu.dk}}}}

\author[2,3,4,5]{Francesco Sannino\footnote{{\Large\orcidlink{0000-0003-2361-5326}}\hspace{0.1cm}\raisebox{0.9ex}{\href{mailto:sannino@qtc.sdu.dk}{sannino@qtc.sdu.dk}}}}

\affil[1]{Universit\'e Claude Bernard Lyon 1, CNRS/IN2P3, IP2I Lyon, UMR 5822, Villeurbanne, F-69100, France}

\affil[2]{Quantum Theory Center (${\hbar}$QTC) \& D-IAS, IMADA at Southern Denmark Univ., Campusvej 55, 5230 Odense M, Denmark}

\affil[3]{Scuola Superiore Meridionale, Largo S. Marcellino, 10, 80138 Napoli NA, Italy}
\affil[4]{Dept. of Physics E. Pancini, Universit`a di Napoli Federico II, via Cintia, 80126 Napoli, Italy}
\affil[5]{INFN sezione di Napoli, via Cintia, 80126 Napoli, Italy}

\begin{document}
\maketitle
\begin{abstract}
 We develop an effective metric description of 2+1 dimensional black holes describing deviations from the classical Bañados--Teitelboim--Zanelli (BTZ) black hole. The latter is a classical 2+1 dimensional rotating black hole with constant negative curvature. The effective metric is constrained by imposing the black hole symmetries and asymptotic classical behavior. The deformed metric is parametrized in terms of a physical quantity that we choose to be a physical distance. The latter can be solved for in three main regions of interest, the one around the horizon, origin, and spatial infinity. The finiteness of physical quantities at the horizon, such as the Ricci and Kretschmann scalars, leads to universal constraints on the physical parameters of the metric around the horizon. This allows us to further derive the general form of the corrected Hawking temperature in terms of the physical parameters of the effective metric. Assuming that the approach can be generalized to the interior of the black hole, we further develop an effective metric description near the origin. To illustrate the approach, we show how to recast the information encoded in a specific model of quantum BTZ known as quBTZ black hole in terms of the effective metric coefficients. 
\end{abstract}
\newpage

\section{Introduction}\label{Sec:1}
Classical general relativity (GR) is a pillar in our understanding of gravitational effects. In fact, it has even predicted the existence of black holes and their classical properties \cite{Schwarzschild:1916uq,Hawking:1976de,Will:2014kxa,Bambi:2019xzp,Jia:2023sef}. Although semiclassical computations have revealed interesting properties beyond classical GR, it is generally agreed that a quantum theory of gravity is still missing \cite{Bertolami:2011aa, Penrose:1964wq, Kiefer:2005uk}. This is, for example, reflected in the fact that GR fails to properly account for the behavior of black holes in the high-energy regime. In view of a lack of such a complete theory of quantum gravity, a plethora of model extensions of GR have been proposed \cite{Lan:2023cvz,Bjerrum-Bohr:2002fji,Cadoni:2022chn}. In particular, within these models, but not only, space-time geometries resembling black holes have been discussed in the literature \cite{bardeen1968,Dymnikova:1992ux,Dymnikova:2004qg,Hayward:2005gi,Frolov:2016pav,Simpson:2018tsi,Simpson:2019mud,Garcia95}.  

\par 
Recently, a new, useful framework has been developed that allows for model independent investigations of black hole metric deviations beyond classical GR \cite{Binetti:2022xdi, DelPiano:2023fiw,DelPiano:2024gvw,DelPiano:2024nrl} (see also \cite{DAlise:2023hls}), notably also including quantum corrections. These new {\it effective metric descriptions} (EMDs) account for metric corrections in terms of physical quantities. This offers the advantage of manifest invariance under coordinate transformations, thereby leading to a universal description encompassing different models.

The original approach focused on four dimensional static, spherically symmetric and asymptotically flat black holes. In this work we generalize the framework to three dimensional stationary black holes with non-vanishing angular momentum that asymptotically approach an AdS space. These can be viewed as deformations of the classical  Bañados--Teitelboim--Zanelli (BTZ) black hole \cite{Banados:1992wn, Banados:1992gq, Bambi:2023try}. Indeed, studying deformations of this particular geometry is interesting for a number of reasons such as: 
\begin{itemize} 
\item[\emph{i)}] {Lower dimensional black holes are interesting toy models to investigate the impact of quantum gravity \cite{Witten:1988hc, Carlip:1995zj}. This opens the possibility of using the EMD framework to effectively bridge among different models while singling out universal features.  }
\item[\emph{ii)}] { Even in the presence of angular momentum three dimensional black hole solutions retain circular symmetry. This might lend some insight on how the four dimensional EMDs can be generalized beyond the static limit. Similarly the inclusion of the cosmological constant in three dimensions might teach us how to take it into account in the four dimensional case. }
\item[\emph{iii)}] {Since asymptotically the BTZ black hole is an ${\text{AdS}_3}$ space with constant, negative curvature it has a dual description in terms of a two dimensional CFT   \cite{Maldacena:1997re,Strominger:1997eq,Witten:1998qj,Carlip:2005zn}. More generally, following the work by Brown and Henneaux \cite{Brown:1986nw}, any consistent quantum theory of gravity on AdS$_3$ displays a conformal symmetry.  Therefore, establishing a universal framework to account for deformations of the classical BTZ metric, preserving the asymptotic AdS geometry, are expected to describe deformations of the dual two dimensional conformal field theory.   }
\end{itemize}
  
We shall, therefore,  develop an effective metric description framework for 2+1D black hole metrics, that classically reduce to the BTZ black hole \cite{Banados:1992wn, Banados:1992gq, Bambi:2023try}. We concentrate on three relevant patches of space-time corresponding to the region near the horizon, near the origin, and asymptotically far away. Imposing finiteness of physical quantities, notably the Ricci and Kretschmann curvature scalars,  we uncover a number of constraints on the physical coefficients of the metric deformation. These coefficients appear when expanding the effective metric deformation in a suitable physical distance. Additionally, we determine the general expression of the Hawking temperature \cite{Hawking:1974rv,Hawking:1976de} in terms of the same physical deformation parameters. We further illustrate the framework, by recasting the information encoded in a specific model of quantum BTZ (quBTZ) constructed through holography \cite{Emparan:2020znc, Panella:2024sor}, in terms of the effective metric coefficients. 
\par

The paper is structured as follows. In Section~\ref{Sec:2} we introduce the effective metric description framework for 2+1D black holes. We then derive the finiteness conditions for the physical coefficients of the effective metric deformation and study thermodynamic properties. Section~\ref{exampl} illustrates how the framework can encode the quBTZ black hole geometry \cite{Emparan:2020znc, Panella:2024sor} via the physical coefficients of the effective metric. Section~\ref{Sec:3} provides a short summary of the presented results and an outlook for future research. Further details of our computations are summarized in the appendices.

\section{Deformed BTZ  Black Holes}\label{Sec:2}
Here we introduce an effective metric description for deformed BTZ black holes generalizing the framework developed in \cite{DelPiano:2023fiw} to two plus one dimensions. 

We start by recalling the metric for the BTZ black hole which reads: 
\begin{equation}\label{2.1}
ds^{2}=-g(r)dt^2+\frac{dr^2}{f(r)}+r^2(h(r)dt+d\phi)^{2},
\end{equation}
with \cite{Banados:1992wn}
\begin{equation}\label{2.2}
g(r)=f(r)=-M-r^2\Lambda+\frac{J^2}{4r^2}, \ \ \ \ h(r)=-\frac{J}{2r^2}, \ \ \ \sqrt{-\Lambda}|J|\leq M.
\end{equation}
The physical parameters are the black hole mass $M$, the angular momentum $J$ and the cosmological constant $\Lambda<0$. The latter induces a non-vanishing constant AdS curvature. Although the metric is superficially divergent at the origin, the geometry is well defined.   
We parametrize deformations of the stationary metric (\ref{2.1}), preserving rotational symmetry as follows:
\begin{equation}\label{2.4}
\begin{aligned}
&f(r)=-M-\left(r^2\Lambda-\frac{J^2}{4r^2}\right) \Phi(d(r)), \ \ \ \ g(r)=-M-\left(r^2\Lambda-\frac{J^2}{4r^2}\right)\Psi(d(r)),\\
&h(r)=-\frac{J}{2r^2}\Omega(d(r)).
\end{aligned}
\end{equation}
Following the framework outlined in 
\cite{DelPiano:2024gvw} the newly introduced ${\Phi, \Psi}$ and   ${\Omega}$ are functions  of a physical quantity which we choose to be the physical distance from the origin: 
\begin{equation}\label{2.3}
    d(r)=\int_0^r \frac{dz}{\sqrt{|f(z)|}} \ .
\end{equation}
This guarantees that the resulting metric enjoys the same properties under coordinate transformations as the original BTZ metric (\ref{2.1}).\par
Furthermore, to preserve the asymptotic behavior of \eqref{2.4} we impose
\begin{equation}\label{2.5}
    \lim_{d\to\infty} \Phi(d)= \lim_{d\to\infty}\Psi(d)= \lim_{d\to\infty}\Omega(d)=1.
\end{equation}
We also require the existence of a simple event horizon located at ${r=r_{H}}$\footnote{We do not assume that this is the only horizon, but just that it is the outermost one. All quantities evaluated at ${r=r_H}$ are denoted with a subscript ${H}$. For example, ${\Phi(d(r_H))=\Phi_H}$ and ${\Psi(d(r_H))=\Psi_H}$.} for which 
\begin{equation}\label{2.6}
\Phi_H=\Psi_H=\frac{4Mr_H^2}{J^2-4r_H^4\Lambda}.
\end{equation} 
\begin{figure}
    \centering
    \includegraphics[width=\textwidth]{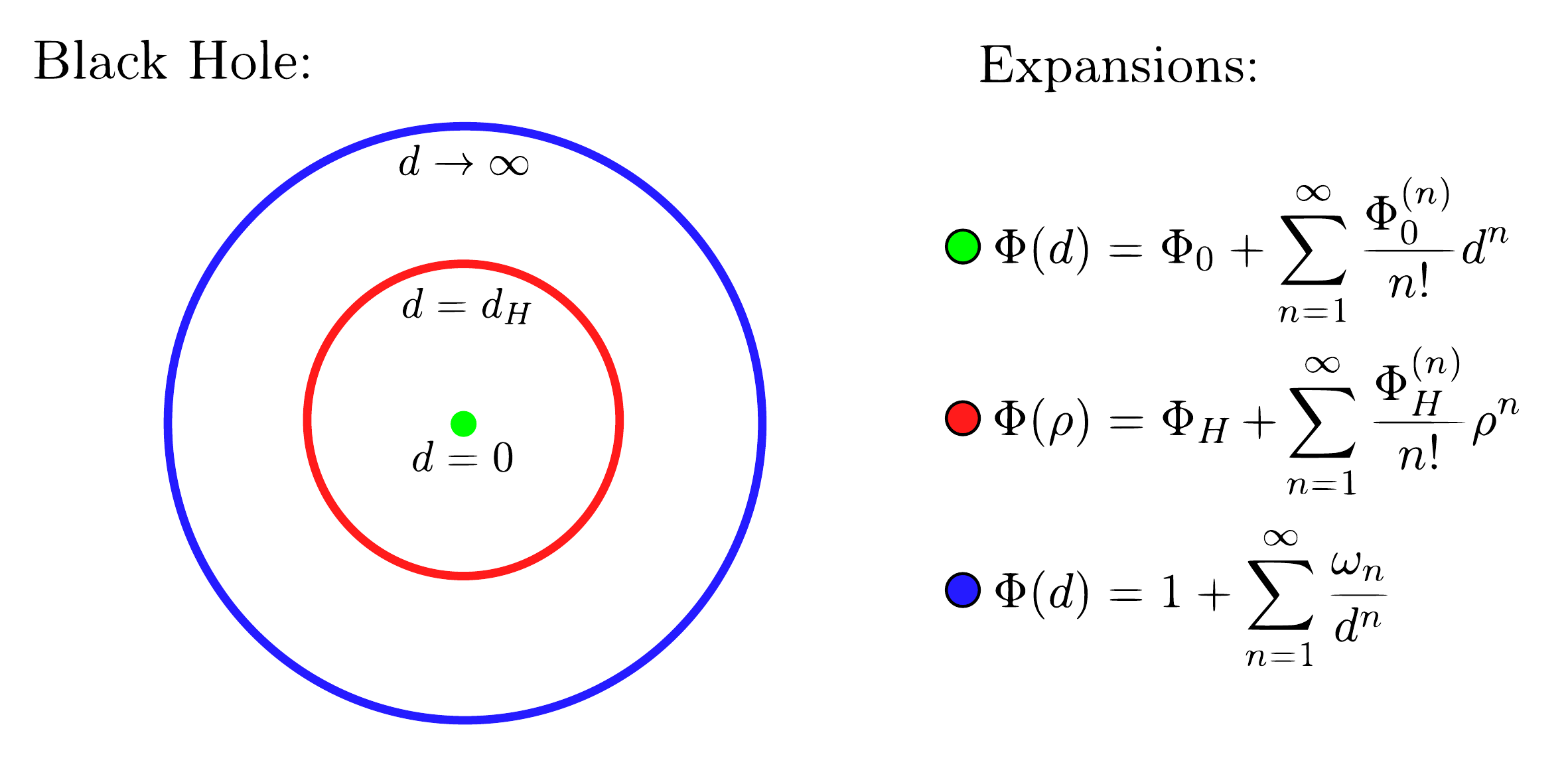}
    \caption{Expansions of the quantum corrected BTZ black hole metric deformation function ${\Phi}$ in different regions. The full metric is expanded in power series of ${\rho}$ close to the outer horizon \tikzcircle[fill=red]{4pt}, and in Taylor series of ${d^{-1}}$ far away from the horizon \tikzcircle[fill=blue]{4pt}. In addition, we also expand the metric in powers of ${d}$ at the origin of the black hole \tikzcircle[fill=green]{4pt}.}\label{Fig.1}
\end{figure}\par
Because equation \eqref{2.3} is a non-linear integral equation for $d$ one can solve for it locally around patches of space-time (see, Fig.~\ref{Fig.1}). The coefficients of the expansion in the physical distance for the functions in \eqref{2.4} are physical because they are coordinate independent. It is natural to explore the following three regions of space-time: 
\begin{itemize}
\item[]{\underline{Near horizon:} A number of physical observables depend on the information accessible in this region, such as the Hawking temperature \cite{Hawking:1974rv}. Furthermore, in the absence of matter sources we expect the curvature to be finite. This yields, as we shall see, constraints on the deformation functions. We expand the functions $\Phi, \Psi$ and $\Omega$   in terms of the physical distance from the horizon $\rho$ as follows: 
\begin{equation}
\label{PCH}
\Phi(\rho) = \Phi_H + \sum_{n=1}^{\infty} \frac{\Phi_H^{(n)} }{n!} \rho^n , \quad \Psi(\rho) = \Psi_H + \sum_{n=1}^{\infty} \frac{\Psi_H^{(n)} }{n!} \rho^n , \quad  \Omega(\rho) = \Omega_H + \sum_{n=1}^{\infty} \frac{\Omega_H^{(n)} }{n!} \rho^n . 
\end{equation}
These will allow us to write the general form of the Hawking temperature in terms of physical parameters for any three dimensional black hole. This space-time region is discussed in Subsection~\ref{sec:sec4.1}.}
\item[]{\underline{Large distance:} Sufficiently far away from the horizon the geometry is expected to approach the classical BTZ limit described by (\ref{2.2}). In this regime, the corrections to the metric in (\ref{2.4}) can be expanded in the inverse of a physical distance. The effective physical coefficients $\omega_n,\gamma_n$ and $\sigma_n$ defined via
\begin{equation} 
\label{PCF}
\Phi(d)=1+\sum_{n=1}^{\infty}\frac{\omega_{n}}{d^{n}}, \ \ \ \ \ \ \ \Psi(d)=1+\sum_{n=1}^{\infty}\frac{\gamma_{n}}{d^{n}}, \ \ \ \ \ \ \Omega(d)=1+\sum_{n=1}^{\infty}\frac{\sigma_{n}}{d^{n}}, 
\end{equation}
can be further constrained by enforcing different versions of the positivity conditions, as detailed in \cite{DAlise:2023hls}. This space-time region is discussed in Subsection~\ref{sec:sec4.2}. }
\item[] {\underline{Near the origin:} This region is expected to be most sensitive to quantum corrections. Nevertheless, we still assume that, at least effectively, a metric description of the form \eqref{2.4} is possible. Here the physical parameters $\Phi_0, \Psi_0$ and $\Omega_0$ are defined via the following expansion around the origin:
\begin{equation}
\label{PCO}
\Phi(d) = \Phi_0 + \sum_{n=1}^{\infty} \frac{\Phi_0^{(n)} }{n!} d^n , \quad \Psi(d) = \Psi_0 + \sum_{n=1}^{\infty} \frac{\Psi_0^{(n)} }{n!} d^n , \quad  \Omega(d) = \Omega_0 + \sum_{n=1}^{\infty} \frac{\Omega_0^{(n)} }{n!} d^n . 
\end{equation}
Such a description permits a model independent investigation of the regularity of the geometry at the heart of the black hole\footnote{Absence of an essential singularity at the origin is not sufficient to guarantee geodesic completeness of space-time. The latter requires more sophisticated studies beyond the scope of this work.}. This space-time region is discussed in Subsection~\ref{sec:sec4.3}.  }
\end{itemize}
We assume implicitly that the series expansions (\ref{PCH}), (\ref{PCF}) and (\ref{PCO}) exist and have finite radius of convergence, such that they correctly represent the metric deformations at least in a finite region of space-time. In the following, we start our exploration by dedicating the next section to the region outside the outermost horizon. 

 \subsection{The deformed BTZ black hole near the horizon}\label{sec:sec4.1}
Near the outer horizon, the metric functions ${f,g,h}$ can be expanded in the form
\begin{equation}\label{4.8}
\begin{aligned}
&f(r)=f^{(1)}_{H}(r-r_H)+\frac{1}{2}f^{(2)}_{H}(r-r_H)^{2}+\mathcal{O}((r-r_H)^{3}),\\
&g(r)=g^{(1)}_{H}(r-r_H)+\frac{1}{2}g^{(2)}_{H}(r-r_H)^{2}+\mathcal{O}((r-r_H)^{3}),\\
&h(r)=h_{H}+h^{(1)}_{H}(r-r_H)+\frac{1}{2}h^{(2)}_{H}(r-r_H)^{2}+\mathcal{O}((r-r_H)^{3}),
\end{aligned}
\end{equation}
such that the coefficients of the expansion are derivatives of metric functions evaluated at ${r_H}$. Conditions \eqref{2.6} impose the vanishing of constant terms in the expansions of ${f}$ and ${g}$.\par
In order to rewrite \eqref{4.8} in terms of a physical observable, we introduce the proper distance from the outer horizon $\rho(r)=d(r)-d_H$. Via \eqref{2.3} we arrive at the following expansion for $\rho$
\begin{equation}
\rho(r)=\frac{2 \sqrt{r-r_H}}{\sqrt{f_H^{(1)}}}-\frac{f_H^{(2)}\left(r-r_H\right)^{3 / 2}}{6\left(f_H^{(1)}\right)^{3 / 2}}+\mathcal{O}\left(\left(r-r_H\right)^{5 / 2}\right),
\end{equation}
which we can invert locally to obtain ${r}$ in terms of ${\rho}$
\begin{equation}\label{2.12}
r(\rho)=r_H+\frac{f_H^{(1)}}{4} \rho^2+\frac{f_H^{(1)} f_H^{(2)}}{96} \rho^4+\mathcal{O}\left(\rho^6\right).
\end{equation}
Once the expression above is substituted into \eqref{4.8}, we obtain the coordinate-independent expansion of the metric functions
\begin{equation}\label{4.11}
\begin{aligned}
&f(r(\rho))=\frac{(f^{(1)}_{H})^2}{4}\rho^2+\frac{(f^{(1)}_{H})^2f^{(2)}_{H}}{24}\rho^4+\mathcal{O}(\rho^6),\\
&g(r(\rho))=\frac{f^{(1)}_{H}g^{(1)}_{H}}{4}\rho^2+\frac{f^{(1)}_{H}\left(3f^{(1)}_{H}g^{(2)}_{H}+f^{(2)}_{H}g^{(1)}_{H}\right)}{96}\rho^4+\mathcal{O}(\rho^{6}),\\
&h(r(\rho))=h_{H}+\frac{f^{(1)}_{H}h^{(1)}_{H}}{4}\rho^2+\frac{f^{(1)}_{H}\left(3f^{(1)}_{H}h^{(2)}_{H}+f^{(2)}_{H}h^{(1)}_{H}\right)}{96}\rho^4+\mathcal{O}(\rho^{6}).
\end{aligned}
\end{equation}\par
We have defined $f^{{(n)}}$ as the $n$-th derivative of the function $f$ with respect to $r$. The first derivatives expanded in terms of $\rho$ then read
\begin{equation}\label{4.17}
\begin{aligned}
&f^{(1)}=\frac{\left(J^2-4 r_H^4 \Lambda\right) \Phi^{(1)}_H}{2r_H^2 f_{H}^{(1)}}\frac{1}{\rho}-\frac{\left(J^2+4 r_H^4 \Lambda\right)\Phi_H}{2r_H^3}+\frac{\left(J^2 -4 r_H^4 \Lambda \right) \Phi_H^{(2)}}{2r_H^2f_H^{(1)}}+\mathcal{O}(\rho),\\
&g^{(1)}=\frac{\left(J^2-4 r_H^4 \Lambda\right) \Psi^{(1)}_H}{2r_H^2 f_{H}^{(1)}}\frac{1}{\rho}-\frac{\left(J^2+4 r_H^4 \Lambda\right)\Psi_H}{2r_H^3}+\frac{\left(J^2 -4 r_H^4 \Lambda \right)\Psi_H^{(2)}}{2r_H^2f_H^{(1)}}+\mathcal{O}(\rho),\\
&h^{(1)}=-\frac{J\Omega^{(1)}_H}{r_H^2f_H^{(1)}}\frac{1}{\rho}+\frac{J\Omega_H}{r_H^3}-\frac{J\Omega_H^{(2)}}{r_H^2 f_H^{(1)}}+\mathcal{O}(\rho).
\end{aligned}
\end{equation}
Recalling that $\Lambda$ is negative, the coefficients of the divergent terms in $1/\rho$ cannot vanish unless 
\begin{equation}\label{4.18}
\Phi_H^{(1)}=\Psi_H^{(1)}=\Omega_H^{(1)}=0.
\end{equation} 
 At the horizon one can solve for $f^{(1)}_H$ for which the positive solution\footnote{Here, we choose the root of the equation that reproduces the correct classical limit.} reads: 
\begin{equation}\label{4.19}
f_H^{(1)}=\frac{M}{r_H}\left(\vartheta_H+\varkappa\right), \ \ \ \ \varkappa=\sqrt{\vartheta_H^2+\frac{2r_H^2}{M}\frac{\Phi_H^{(2)}}{\Phi_H}} \ ,
\end{equation}
with $\vartheta_H$ the following function  
\begin{equation}\label{new4.16}
 \vartheta(r)=\frac{4 r^4 \Lambda+J^2}{4 r^4 \Lambda-J^2} \ ,
\end{equation}
evaluated at $r_H$. In the non-rotating limit, ${\vartheta\rightarrow 1}$.\par
The remaining two derivatives become
\begin{equation}\label{4.20}
g_H^{(1)}=\frac{2M}{r_H}\vartheta_H+\frac{2 M}{ f_{H}^{(1)}}\frac{\Psi_H^{(2)}}{\Psi_H}, \ \ \ \ \ h_H^{(1)}=\frac{J\Omega_H}{r_H^3}-\frac{J\Omega_H^{(2)}}{r_H^2 f_H^{(1)}}.
\end{equation}
Imposing reality and positivity of the first derivatives at the horizon bounds also the second derivatives ${\Phi_H^{(2)}}$,${\Psi_H^{(2)}}$  as follows
\begin{equation}\label{eqineq}
\frac{\Phi_H^{(2)}}{\Phi_H}\geq-\frac{M}{2r_H^2}\vartheta_H^2, \ \ \ \frac{\Psi_H^{(2)}}{\Psi_H}>-\frac{M} {r_H^2}\vartheta_H\left(\varkappa+\vartheta_H\right), \ \ \ \vartheta_H>-\varkappa.
\end{equation}
Similarly, imposing finiteness of the second order derivatives at the horizon requires:  
\begin{equation}\label{cond4.26}
\Phi^{(3)}_H=\Psi^{(3)}_H=\Omega^{(3)}_H=0.
\end{equation}
  The above relations, together with conditions \eqref{4.18} are sufficient to ensure finiteness of Ricci and Kretschmann scalars at the horizon. This concludes the discussion of the effective metric description near the outer horizon, with some of the consequences for thermodynamics discussed in Sec.~\ref{sec:sec4.4}.

\subsection{Large distance expansion}\label{sec:sec4.2}
Here we shall explore the large distance expansion starting with expanding the functions ${\Phi, \Psi, \Omega}$ as  follows
\begin{equation}\label{ldexp}
\Phi(d)=1+\sum_{n=1}^{\infty}\frac{\omega_{n}}{d^{n}}, \ \ \ \ \ \ \ \Psi(d)=1+\sum_{n=1}^{\infty}\frac{\gamma_{n}}{d^{n}}, \ \ \ \ \ \ \Omega(d)=1+\sum_{n=1}^{\infty}\frac{\sigma_{n}}{d^{n}}. 
\end{equation}
This expansion, by construction, satisfies the asymptotically AdS boundary conditions \eqref{2.5} at infinity, \emph{i.e.} in the limit $d\to\infty$. If all three series are convergent all the way up to the outer horizon
\begin{equation}\label{4.27}
\lim_{n\to\infty}\text{sup}|\omega_{n}|^{\frac{1}{n}}\leq d_{H}, \ \ \ \ \ \ \ \ \lim_{n\to\infty}\text{sup}|\gamma_{n}|^{\frac{1}{n}}\leq d_{H}, \ \ \ \ \ \ \ \ 
\lim_{n\to\infty}\text{sup}|\sigma_{n}|^{\frac{1}{n}}\leq d_{H},
\end{equation}
then one may use horizon constraints to impose conditions on the rescaled expansion coefficients ${\bar{\omega}_n,\bar{\gamma}_n,\bar{\sigma}_n}$. These coefficients are rescaled by a factor of ${d_H^{-n}}$ with respect to ${{\omega}_n,{\gamma}_n,{\sigma}_n}$.\par
The position of the outer horizon ${r_{H}}$ is given by \eqref{2.6}
\begin{equation}\label{4.28}
\sum_{n=1}^{\infty}\bar{\omega}_n=\sum_{n=1}^{\infty}\bar{\gamma}_{n}=\frac{4Mr_H^2}{J^2-4r_H^4\Lambda}-1.
\end{equation}
As in the near horizon expansion, we can rewrite derivatives of ${f,g,h}$ in terms of the proper distance, and isolate divergent terms. For the first derivatives to converge at the horizon, we require:
\begin{equation}\label{cond1}
\sum_{n=1}^{\infty}n\bar{\omega}_{n}=\sum_{n=1}^{\infty}n\bar{\gamma}_{n}=\sum_{n=1}^{\infty}n\bar{\sigma}_{n}=0.
\end{equation}
With those relations imposed, we find the expressions for the first derivatives of the metric functions:
\begin{equation}
\begin{aligned}
&f_H^{(1)}=\frac{M}{r_H}\left(\vartheta_H+\bar{\varkappa}\right), \ \ \ \bar{\varkappa}=\sqrt{\vartheta_H^2+\frac{J^2-4r_H^4\Lambda}{2d_H^2 M^2}\sum_{n=1}^{\infty}n^2\bar{\omega}_{n}},\\
&g_H^{(1)}=\frac{2M}{r_H}\vartheta_H+\frac{J^2-4r_H^4\Lambda}{2d_H^2r_H^2f_H^{(1)}}\sum_{n=1}^{\infty}n^2\bar{\gamma}_{n},\\
&h_H^{(1)}=-\frac{2h_H}{r_H}-\frac{J}{d_H^2r_H^2f_H^{(1)}}\sum_{n=1}^{\infty}n^2\bar{\sigma}_{n}.
\end{aligned}
\end{equation}
The above expressions can be compared with near horizon expansions (\ref{4.19}-\ref{4.20}) to identify
\begin{equation}
\begin{aligned}
&\Phi_H^{(2)}=\frac{1}{d_H^2}\sum_{n=1}^{\infty}n^2\bar{\omega}_{n}, \ \ \ \Psi_H^{(2)}=\frac{1}{d_H^2}\sum_{n=1}^{\infty}n^2\bar{\gamma}_{n}, \ \ \ \Omega_H^{(2)}=\frac{1}{d_H^2}\sum_{n=1}^{\infty}n^2\bar{\sigma}_{n}. 
\end{aligned}
\end{equation}\par
Finally, we can investigate the second derivative constraints. In terms of large distance coefficients, the regularization conditions turn out to be
\begin{equation}\label{cond2}
\sum_{n=1}^{\infty} n^2(n+3) \bar{\omega}_n=\sum_{n=1}^{\infty} n^2(n+3) \bar{\gamma}_n=\sum_{n=1}^{\infty} n^2(n+3) \bar{\sigma}_n=0.
\end{equation}
Conditions \eqref{cond1} and \eqref{cond2} are sufficient for the Ricci and Kretschmann scalars to remain finite at the horizon.
\subsection{Expansion at the origin}\label{sec:sec4.3}
In contrast to other black hole solutions (\emph{e.g.} the Schwarzschild metric \cite{Schwarzschild:1916uq} in 4 dimensions), the BTZ black hole is well behaved at ${r=0}$, i.e. there is no essential singularity. The apparent metric singularity for ${J\neq 0}$ is due to the choice of polar coordinates. Assuming that a complete theory of quantum gravity would still generate a well defined metric at the origin, we can expand the metric in equation \eqref{2.4} in powers of the proper distance and examine the finiteness of the scalar invariants. In order to do so, we assume that the correction functions ${\Phi,\Psi,\Omega}$ are analytical at the origin. As a consequence of this assumption, we have shown in Appx.~\ref{Appx:A} that the expansion of ${f,g,h}$ in terms of ${r}$ assumes the form
\begin{equation}\label{4.37}
\begin{aligned}
&f(r)=f_{0}^{(-2)}\frac{1}{r^2}+f_{0}+f^{(2)}_{0}r^{2}+\mathcal{O}(r^{3}),\\
&g(r)=g_{0}^{(-2)}\frac{1}{r^2}+g_{0}+g^{(2)}_{0}r^{2}+\mathcal{O}(r^{3}),\\
&h(r)=h_{0}^{(-2)}\frac{1}{r^2}+h_{0}+h^{(2)}_{0}r^{2}+\mathcal{O}(r^{3}).
\end{aligned}
\end{equation}\par
In Sec.~\ref{exampl} we shall see, we will encounter an example that does not follow \eqref{4.37} and in fact has an essential singularity at the origin. Using the first equation of \eqref{4.37}, and assuming a non-vanishing $f_{0}^{(-2)}$, we can expand the proper distance \eqref{2.3} near the origin
\begin{equation}\label{proporigin}
d (r)=\int_{0}^{r} \frac{zdz}{\sqrt{\left|f_{0}^{(-2)}+f_{0}z^2+f^{(2)}_{0}z^{4}+...\right|}}=\frac{r^2}{2 \sqrt{\left|f_{0}^{(-2)}\right|}}-\frac{f_{0}}{8 f_{0}^{(-2)}\sqrt{\left|f_{0}^{(-2)}\right|}}r^4+\mathcal{O}\left(r^5\right),
\end{equation} 
and invert it locally to write the radial coordinate as power series in ${d}$:
\begin{equation}\label{4.41}
r(d)=\sqrt{2}\left|f_0^{(-2)}\right|^{1 / 4}d^{1/2}+\frac{ f_0\left|f_0^{(-2)}\right|^{3/4}}{2 \sqrt{2} f_0^{(-2)}}d^{3 / 2}+\mathcal{O}(d^{5/2}).
\end{equation}
To extract the physical coefficients, we compare the following expansion in terms of ${d}$, which can be obtained from substituting \eqref{4.41} into \eqref{4.37}:
\begin{equation}
\begin{aligned}
f=&\frac{f_{0}^{(-2)}}{2\sqrt{\left|f_{0}^{(-2)}\right|} d}+\frac{3 f_0}{4}+\frac{\left(3 f_0^2+40 f_0^{(-2)} f_0^{(2)} \right)f_0^{(-2)}}{24 \left|f_0^{(-2)}\right|^{3/2}}d+\mathcal{O}(d^{2}),\\
g=&\frac{g_0^{(-2)}}{2 \sqrt{\left|f_0^{(-2)}\right|} d}+g_0-\frac{f_{0} g_0^{(-2)}}{4 f_0^{(-2)}}+\frac{\left(3 f_0^{2} g_0^{(-2)}-8 f_0^{(-2)} f_0^{(2)} g_0^{(-2)}+48 \left(f_0^{(-2)}\right)^2 g_0^{(2)}\right)}{24 \left|f_0^{(-2)}\right|^{3 / 2}}d+\mathcal{O}(d^{2}),\\
h=&\frac{h_0^{(-2)}}{2 \sqrt{\left|f_0^{(-2)}\right|} d}+h_0-\frac{f_{0} h_0^{(-2)}}{4 f_0^{(-2)}}+\frac{\left(3 f_0^{2} h_0^{(-2)}-8 f_0^{(-2)} f_0^{(2)} h_0^{(-2)}+48 \left(f_0^{(-2)}\right)^2 h_0^{(2)}\right)}{24 \left|f_0^{(-2)}\right|^{3 / 2}}d+\mathcal{O}(d^{2}),\\
\end{aligned}
\end{equation}
with the one originating from \eqref{2.4}. Comparing the two expansions order by order allows us to transform between the two parametrizations of ${f,g,h}$ in terms of ${r}$ (with coefficients ${f_0^{(n)}, g_0^{(n)}, h_0^{(n)}}$) and ${d}$ (with coefficients ${\Phi_0^{(n)}, \Psi_0^{(n)}, \Omega_0^{(n)}}$):
\begin{equation}
\begin{aligned}
&\Phi_0=\frac{4 f_0^{(-2)}}{J^2}, \ \ \ \Psi_0=\frac{4g_0^{(-2)}}{J^2},\ \ \ \  \Omega_0=-\frac{2h_0^{(-2)}}{J},\\
&\Phi_0^{(1)}=\frac{8\sqrt{\left|f_0^{(-2)}\right|}}{J^2}\left(M+f_0\right), \ \ \ \Psi_0^{(1)}=\frac{8\sqrt{\left|f_0^{(-2)}\right|}}{J^2}\left(M+g_0\right),\ \ \ \  \Omega_0^{(1)}=-\frac{4\sqrt{\left|f_0^{(-2)}\right|}h_0}{J},\\
&\Phi_0^{(2)}=\frac{8f_0^{(-2)}}{J^4\left|f_0^{(-2)}\right|}\left(16 \Lambda \left(f_0^{(-2)}\right)^2+J^2 M f_0+J^2 f_0^2+4 J^2 f_0^{(-2)} f_0^{(2)}\right),\\
&\Psi_0^{(2)}=\frac{8f_0^{(-2)}}{J^4\left|f_0^{(-2)}\right|}\left(16\Lambda f_0^{(-2)}g_0^{(-2)}+J^2 M f_0+J^2 f_0 g_0+4 J^2 f_0^{(-2)}g_0^{(2)}\right),\\
&\Omega_0^{(2)}=-\frac{4f_0^{(-2)}}{J\left|f_0^{(-2)}\right|}\left(f_0 h_0+4f_0^{(-2)}h_0^{(2)}\right).
\end{aligned}
\end{equation}\par
The finiteness of the Ricci and Kretschmann scalars at the origin then requires
\begin{equation}\label{expzero4.47}
\Omega_0^2=\Psi_0.
\end{equation}
The classical BTZ black hole trivially satisfies the above conditions\footnote{Throughout the expansion, we assumed ${\Phi_0,\Psi_0,J\neq 0}$ for simplicity. Nevertheless, analogous calculations can be carried out if this condition is not satisfied. For the ${J=0}$ case, the condition \eqref{expzero4.47} is always satisfied, but might not be sufficient to guarantee the finiteness of the Ricci and Kretschmann scalars.}.

\subsection{Finiteness conditions summarized}\label{Appx:B}
Here, we summarize all the regularity conditions for the quantum corrected BTZ black hole. These can be categorized into three separate groups according to how they were derived.\par
{
\centering
\begin{center}
{\bf Near Horizon Expansion (conditions at ${r=r_H}$)}
\end{center}

\begin{tcolorbox}
\begin{center}
$\Phi_H^{(1)}=\Psi_H^{(1)}=\Omega_H^{(1)}=0$, \ \ \ $\Phi_H^{(3)}=\Psi_H^{(3)}=\Omega_H^{(3)}=0$, \ \ \ $\Phi_H=\Psi_H=\frac{4Mr_H^2}{J^2-4r_H^4\Lambda}$,\par
\vspace{0.1cm}
$\frac{\Phi_H^{(2)}}{\Phi_H}\geq-\frac{M}{2r_H^2}\vartheta_H^2, \ \ \ \frac{\Psi_H^{(2)}}{\Psi_H}>-\frac{M}{r_H^2}\vartheta_H\left(\varkappa+\vartheta_H\right), \ \ \ \vartheta_H>-\varkappa$

\end{center}
\end{tcolorbox}
\begin{center}
{\bf Large Distance Expansion (conditions at ${r=r_H}$)}
\end{center}

\begin{tcolorbox}
\begin{center}
$\sum_{n=1}^{\infty}n\bar{\omega}_{n}=\sum_{n=1}^{\infty}n\bar{\gamma}_{n}=\sum_{n=1}^{\infty}n\bar{\sigma}_{n}=0$, \ \ \ $\sum_{n=1}^{\infty}\bar{\omega}_n=\sum_{n=1}^{\infty}\bar{\gamma}_{n}=\frac{4Mr_H^2}{J^2-4r_H^4\Lambda}-1$, \par
\vspace{0.2cm}
$\sum_{n=1}^{\infty} n^2(n+3) \bar{\omega}_n=\sum_{n=1}^{\infty} n^2(n+3) \bar{\gamma}_n=\sum_{n=1}^{\infty} n^2(n+3) \bar{\sigma}_n=0$,\par
\vspace{0.2cm}
$\frac{\sum_{n=1}^{\infty}n^2\bar{\omega}_{n}}{1+\sum_{n=1}^{\infty}\bar{\omega}_{n}}\geq-\frac{M d_H^2}{2r_H^2}\vartheta(r_H)^2, \ \ \ \frac{\sum_{n=1}^{\infty}n^2\bar{\gamma}_{n}}{1+\sum_{n=1}^{\infty}\bar{\gamma}_{n}}>-\frac{M d_H^2}{r_H^2}\vartheta_H\left(\bar{\varkappa}+\vartheta_H\right), \ \ \ \vartheta_H>-\bar{\varkappa}$
\vspace{0.2cm}

\end{center}
\end{tcolorbox}

\begin{center}
{\bf Origin Expansion (conditions at ${r=0}$)}
\end{center}

\begin{tcolorbox}
\begin{center}
$\Omega_0^2=\Psi_0$ 
\end{center}
\end{tcolorbox}
}These conditions are expected to hold for different extensions of the BTZ metric stemming from different gravity models. 

\subsection{Thermodynamics}\label{sec:sec4.4}
Having developed the framework for describing BTZ black hole deformations, we now proceed to investigate their thermodynamical properties. The temperature for a rotating 2+1D black hole can be expressed in terms of the metric functions via (see \emph{e.g.} \cite{Wald:1984rg})
\begin{equation}\label{2.54}
\begin{aligned}
T=\frac{\kappa}{2\pi}=\frac{1}{4\pi}\sqrt{f_H^{(1)}g_H^{(1)}},
\end{aligned}
\end{equation}
where ${\kappa}$ is the surface gravity at the horizon.\par
The above equation can be re-expressed in terms of the near horizon and large distance expansions, developed in Sec.~\ref{sec:sec4.1} and Sec.~\ref{sec:sec4.2}, respectively. We substitute the first derivatives into \eqref{2.54} to formulate the black hole temperature in terms of the quantum correction functions ${\Phi, \Psi, \Omega}$
\begin{equation}\label{temp1eq}
T=\frac{1}{2\pi}\sqrt{\frac{M^2}{2 r_H^2}\vartheta_H\left(\vartheta_H+\varkappa\right)+\frac{M}{2}\frac{\Psi_H^{(2)}}{\Psi_H}}.
\end{equation}
If the large distance expansion \eqref{ldexp} converges all the way to the horizon, we can also rewrite the temperature in terms of the large distance expansion coefficients
\begin{equation}\label{temp2eq}
T=\frac{1}{2\pi}\sqrt{\frac{M^2}{2 r_H^2}\vartheta_H\left(\vartheta_H+\bar{\varkappa}\right)+\frac{M}{2d_H^2}\frac{\sum_{n=1}^{\infty}n^2\bar{\gamma}_{n}}{1+\sum_{n=1}^{\infty}\bar{\gamma}_{n}}}.
\end{equation}
We can also verify that in the case of the BTZ solution \eqref{2.2}, temperatures (\ref{temp1eq}-\ref{temp2eq}) reduce to that of a rotating BTZ black hole
\begin{equation}
T=-\frac{\Lambda}{2\pi}\frac{r_{+}^2-r_{-}^2}{r_{+}}, \ \ \ r_{ \pm}^2=-\frac{M}{2\Lambda}\left[1 \pm\left(1+\frac{J^2\Lambda}{M^2}\right)^{1 / 2}\right],
\end{equation}
in agreement with the classical result derived in \cite{Carlip:1995qv}.\par
The first law of black hole thermodynamics relating the temperature, entropy $S$ and angular momentum states that:
\begin{equation}\label{2.40}
TdS=dM-\omega_H dJ, \ \ \ \omega_H=-h_H=\frac{J}{2r_H^2}\Omega_H.
\end{equation}
To obtain the entropy, one should integrate the first law which, however, implies knowledge of the dependence upon ${M,J}$ of the effective coefficients ${h_H,r_H,\varkappa,\vartheta_H,\frac{\Psi_H^{(2)}}{\Psi_H}}$
\begin{equation}\label{2.40int}
S(M,J)=\int dS=\int_{0,0}^{M,J} \frac{1}{T}\left(dM-\omega_H dJ\right).
\end{equation}
Assuming that the previous expression can be integrated, one would arrive at the entropy as a function of ${M}$ and ${J}$. This can be performed, however, only if the integral is well defined along the chosen path of integration\footnote{In fact, one would even require that the result \eqref{2.40int} is independent of a choice of this path.}. This is not always guaranteed, and it will have to be checked model by model. 

\section{Holographic quBTZ Black Hole}\label{exampl}
As an example, we apply the effective metric description to a quantum corrected BTZ black hole, constructed through braneworld holography \cite{Emparan:2020znc, Emparan:2021hyr}. For simplicity, we restrict ourselves to considering the non-rotating quBTZ black hole, whose metric reads
\begin{equation}\label{metricquBTZ}
d s^2=-\left(\frac{r^2}{\ell_3^2}-F_1-\frac{\ell F_2}{r}\right) d t^2+\frac{d r^2}{\frac{r^2}{\ell_3^2}-F_1-\frac{\ell F_2}{r}}+r^2 d \phi^2.
\end{equation}
The metric is fully fixed by specifying the black hole mass parameter ${z}$, backreaction parameter ${\nu}$ and the $\text{AdS}_3$ radius on the brane ${\ell_3}$ via\footnote{The parameter ${\ell}$ is the inverse of brane tension. It appears in the effective action as a cut-off length for the 3D effective field theory. Physically, it quantifies the strength of the backreaction of matter fields in the resulting effective theory, and in turn the degree to which the metric deviates from that of the classical BTZ black hole.}
\begin{equation}\label{eq4.61}
F_1=4\frac{z^2\left(1-\nu z^3\right)(1+\nu z)}{\left(1+3 z^2+2 \nu z^3\right)^2}, \ \ \ F_2=8 \frac{z^4\left(1+z^2\right)(1+\nu z)^2}{\left(1+3 z^2+2 \nu z^3\right)^3}, \ \ \ \ell=\nu \ell_3,
\end{equation} 
where the metric functions ${f, g}$ read
\begin{equation}\label{4.56}
g(r)=f(r)=\frac{r^2}{\ell_3^2}-F_1-\frac{\ell F_2}{r}.
\end{equation}
We note that \eqref{4.56} contains an additional ${r^{-1}}$ term not present in the expansion \eqref{4.37}. Therefore, based on the result from Appx.~\ref{Appx:A}, we can conclude that the correction functions are not analytic around the origin. In fact, the metric (\ref{metricquBTZ}) suffers from an essential singularity at ${r=0}$ \cite{Emparan:2020znc}.\par
At the horizon, the metric is well-behaved and we can calculate the temperature of the quBTZ black hole. The position of the outer horizon is a solution to the equation
\begin{equation}
\frac{r_H^2}{\ell_3^2}-F_1-\frac{\ell F_2}{r_H}=0 \implies r_H^2=\frac{4l_3^2 z^2(1+\nu z)^2}{(1+3z^2+2\nu z^3)^2}.
\end{equation}
Comparing \eqref{4.56} with \eqref{2.4}, we obtain
\begin{equation}\label{4.62}
\begin{aligned}
\Phi(d(r))=\Psi(d(r))=\frac{F_1-M}{\Lambda r^2}+\frac{\ell F_2}{\Lambda r^3}-\frac{1}{\Lambda \ell_3^2}.
\end{aligned}
\end{equation}
In the limit of infinite brane tension ${\ell\rightarrow 0}$, the metric reduces to the classical BTZ solution, for which ${\Phi(d(r)),\Psi(d(r))=1}$:
\begin{equation}\label{eq4.63}
\begin{aligned}
\forall r>0, \ \ \ \frac{F_1-M}{\Lambda r^2}-\frac{1}{\Lambda \ell_3^2}=1.
\end{aligned}
\end{equation}
This allows us to further identify ${M\equiv F_1}$, ${\Lambda\equiv -1/\ell_3^2}$, i.e.
\begin{equation}\label{4.63}
\begin{aligned}
\Phi(d(r))=\Psi(d(r))=1+\frac{\ell F_2}{\Lambda r^3}=1-8 \frac{\nu \ell_3^3 z^4\left(1+z^2\right)(1+\nu z)^2}{r^3\left(1+3 z^2+2 \nu z^3\right)^3}.
\end{aligned}
\end{equation}
When expanded in ${d}$, the first and third derivatives of ${\Phi, \Psi}$ indeed vanish at the horizon, which ensures the finitness of the Ricci and Kretschmann scalars. By substituting \eqref{4.63} to \eqref{temp1eq}, the temperature of the black hole \eqref{temp1eq} can be then cast into the form
\begin{equation}
T=\frac{M}{4\pi r_H}\left(1+\sqrt{1+\frac{2r_H^2}{M}\frac{\Phi_H^{(2)}}{\Phi_H}}\right)=\frac{z}{2 \pi \ell_3} \frac{2+3 \nu z+\nu z^3}{1+3 z^2+2 \nu z^3},
\end{equation}
which agrees with the result found in \cite{Emparan:2020znc}.

\section{Conclusions and Further Considerations}\label{Sec:3}

In this work we extended the effective metric approach developed in \cite{Binetti:2022xdi, DelPiano:2023fiw} to 2+1 dimensional black holes describing deviations from the Bañados--Teitelboim--Zanelli (BTZ) black hole \cite{Banados:1992gq}. To constrain the metric coefficients, we utilized the asymptotic behavior and space-time symmetries. Furthermore, the finiteness of physical quantities at the horizon, such as the Ricci and Kretschmann scalars, leads to universal constraints on the physical coefficients in \eqref{PCH}. The approach allowed us to then determine the general form of the Hawking temperature in terms of those physical coefficients. Assuming that the approach can be generalized to the interior of the black hole, we developed an effective metric description near the origin. We then illustrate the approach via the example of the quBTZ black hole \cite{Emparan:2020znc, Emparan:2021hyr}. 
We have shown that it is possible to construct an effective metric description for three dimensional black holes that naturally encompasses a wide class of deformed BTZ metrics. An important feature of our approach is that the metric deviations are encoded in a model independent way, with the specific underlying models captured by identifying the given values of the physical metric coefficients. 

Since the classical BTZ black hole is an ${\text{AdS}_3}$ space with constant, negative curvature, it has been employed within the context of the AdS-CFT correspondence \cite{Strominger:1997eq,Witten:1998qj,Carlip:2005zn}. We therefore plan to apply our framework to generalize the AdS-CFT correspondence beyond the BTZ limit, which is expected to inform us on the allowed deformations of the dual CFT.

\subsection*{Acknowledgments}
We thank Manuel Del Piano, Vania Vellucci, and John Wheater for enlightening discussions on several related topics. S.H. would like to thank the Quantum Theory Center (QTC) at the Danish Institute for Advanced Study and IMADA of the University of Southern Denmark for the hospitality during the completion of this work. The work of F.S. is partially supported by the Carlsberg Foundation, grant CF22-0922.

\appendix

 \section{General Form of the Expansion at the Origin}\label{Appx:A}
In this Appendix, we show that when correction functions ${\Phi,\Psi,\Omega}$ are analytic near the origin, \eqref{4.37} is the most general expansion of the metric functions near ${r=0}$. Suppose for now that the metric function $f$ admits an expansion at the origin in the form of generic Laurent series
\begin{equation}\label{App1}
f(r)=r^{-k}\sum_{n=0}^{\infty}c_n r^n,
\end{equation}
of order $k\in Z_{+}$. We assume that this is a formal Laurent series, i.e. that there is only a finite number of terms with negative power of ${r}$. This ensures that the multiplication of the series of the form \eqref{App1} is well defined.\par
The radial proper distance \eqref{2.3} then reads:
\begin{equation}
\begin{split}
&d (r)=\int_{0}^{r} \frac{z^{k/2} dz}{\sqrt{\left|c_0+c_1 z+c_2 z^2+c_3 z^3+c_4 z^{4}+...\right|}}=\\
&=\frac{2 r^{1+\frac{k}{2}}}{(2+k) \sqrt{|c_0|}}-\frac{c_1 r^{2+\frac{k}{2}}}{(4+k) c_0 \sqrt{|c_0|}}+\frac{3 c_1^2-4 c_0 c_2 }{4 (6+k)c_0^{2}\sqrt{|c_0|}}r^{3+\frac{k}{2}}+\mathcal{O}(r^{4+\frac{k}{2}}).
\end{split}
\end{equation} 
The above relation can be locally inverted
\begin{equation}\label{propdistapp}
r(d)=|c_0|^{\frac{1}{k+2}}\left(\frac{k+2}{2}\right)^{\frac{2}{k+2}}d^{\frac{2}{k+2}}+\frac{c_1 |c_0|^{\frac{2}{k+2}}\left(\frac{k+2}{2}\right)^{\frac{4}{k+2}}}{c_0(k+4)}d^{\frac{4}{k+2}}+\mathcal{O}\left(d^{\frac{6}{k+2}}\right),
\end{equation}
and substituted back to \eqref{App1} to obtain the expansion of ${f}$ in the proper distance around the origin. Although the expansion coefficients are complex, non-linear combinations of ${c_n}$, we know that the expansion takes the form
\begin{equation}\label{Appx79}
f(r)=\sum_{n=0}^{\infty} \tilde{c}_n d^{\frac{2n-2k}{k+2}}.
\end{equation}
Alternatively, we can also expand the function $f$ using the parametrization \eqref{2.4}. Since ${\Phi}$ can be Taylor expanded around the origin, the general form of the resulting series reads
\begin{equation}\label{Appx80}
f(r)=\sum_{n_1=0}^{\infty}\sum_{n_2=0}^{\infty}\tilde{b}_{n_1}\tilde{e}_{n_2} d^{2\frac{n_1-2}{k+2}+n_2}.
\end{equation}
Since \eqref{Appx79} and \eqref{Appx80} are two expansions of the same function, they have to coincidence. In particular, the powers of ${d}$ in both expansions have to coincidence, which imposes
\begin{equation}
    \frac{-2k}{k+2}=\frac{-4}{k+2} \implies k=2.
\end{equation}
Therefore, the two expansions can overlap only if ${k=2}$, i.e. ${f}$ takes the form 
\begin{equation}\label{Appx85}
    f(r)=f_{0}^{(-2)}\frac{1}{r^2}+f_{0}^{(-1)}\frac{1}{r}+f_{0}+f^{(1)}_{0}r+f^{(2)}_{0}r^{2}+\mathcal{O}(r^{3}).
\end{equation}
\par
Finally, we can use \eqref{Appx85} to calculate the proper distance \eqref{propdistapp} and expand ${f}$ in the radial coordinate ${r}$ using the original parametrization \eqref{2.4}. The obtained expression contains only even powers of ${r}$:
\begin{equation}
\begin{aligned}
&f(r)=\frac{J^2 \Phi_0}{4 r^2}-\left(M-\frac{J^2 \Phi_0^{(1)}}{8 \sqrt{\left|f_0^{(-2)}\right|}}\right)-\left(\Lambda \Phi_0+\frac{J^2 f_0 \Phi^{(1)}_0}{32 f_0^{(-2)} \sqrt{\left|f_0^{(-2)}\right|}}-\frac{J^2\Phi^{(2)}_0}{32 \left|f_0^{(-2)}\right|}\right) r^2+\mathcal{O}(r^4).
\end{aligned}
\end{equation}
This further narrows down the possible form of expansion of ${f}$ to
\begin{equation}\label{eqA87}
    f(r)=f_{0}^{(-2)}\frac{1}{r^2}+f_{0}+f^{(2)}_{0}r^{2}+\mathcal{O}(r^{4}).
\end{equation}
As a result, \eqref{eqA87} is the most general formal Laurent series of $f$ for which ${\Phi}$ can be analytic around the origin. Analogous analysis can be carried out for ${g}$ and ${h}$, leading to the two expansions at the bottom of equation \eqref{4.37}.\par

\section{Ricci and Kretschmann Scalars}\label{Appx:C}
Below, we give the expressions for the Ricci and Kretschmann invariants used throughout the paper. The invariants are calculated for the metric \eqref{2.1}.\par
\centering
\vspace{0.4cm}
{\bf Ricci Scalar:}
\begin{align*}
R=-\frac{f^{(1)}}{r}-\frac{f g^{(1)}}{r g}-\frac{f^{(1)} g^{(1)}}{2 g}+\frac{f (g^{(1)})^2}{2 g^2}+\frac{r^2 f (h^{(1)})^2}{2 g}-\frac{f g^{(2)}}{g}
\end{align*}

{\bf Kretschmann Scalar:}
\begin{align*}
&K= \frac{(f^{(1)})^2}{r^2}+\frac{f^2 (g^{(1)})^2}{r^2 g^2}+\frac{(f^{(1)})^2 (g^{(1)})^2}{4 g^2}-\frac{f f^{(1)} (g^{(1)})^3}{2 g^3}+\frac{f^2 (g^{(1)})^4}{4 g^4}-\frac{18 f^2 (h^{(1)})^2}{g} \\
& -\frac{5 r f f^{(1)} (h^{(1)})^2}{g}-\frac{r^2 (f^{(1)})^2 (h^{(1)})^2}{2 g}+\frac{7 r f^2 g^{(1)} (h^{(1)})^2}{g^2}-\frac{r^2 f f^{(1)} g^{(1)} (h^{(1)})^2}{2 g^2}\\
&+\frac{r^2 f^2 (g^{(1)})^2 (h^{(1)})^2}{g^3}+\frac{11 r^4 f^2 (h^{(1)})^4}{4 g^2}+\frac{f f^{(1)} g^{(1)} g^{(2)}}{g^2}-\frac{f^2 (g^{(1)})^2 g^{(2)}}{g^3}\\
&-\frac{3 r^2 f^2 (h^{(1)})^2 g^{(2)}}{g^2}+\frac{f^2 (g^{(2)})^2}{g^2}-\frac{12 r f^2 h^{(1)} h^{(2)}}{g}-\frac{2 r^2 f f^{(1)} h^{(1)} h^{(2)}}{g}\\
&+\frac{2 r^2 f^2 g^{(1)} h^{(1)}h^{(2)}}{g^2}-\frac{2 r^2 f^2 (h^{(2)})^2}{g}
\end{align*}

\sloppy

\newpage
\printbibliography[heading=bibintoc,title={References}]
\end{document}